# A Task-Driven Human-AI Collaboration:

## When to Automate, When to Collaborate, When to Challenge

Saleh Afroogh
Urban Information Lab
University of Texas at Austin
saleh.afroogh@utexas.edu

Kush R. Varshney
IBM Research
Yorktown Heights, USA
krvarshn@us.ibm.com

Jason D'Cruz
State University of New York
at Albany, USA
jdcruz@albany.edu

**Abstract**

According to several empirical investigations, despite enhancing human capabilities, human-AI cooperation frequently falls short of expectations and fails to reach true synergy. We propose a task-driven framework that reverses prevalent approaches by assigning AI roles according to how the task's requirements align with the capabilities of AI technology. Three major AI roles are identified through task analysis across risk and complexity dimensions: autonomous, assistive/collaborative, and adversarial. We show how proper human-AI integration maintains meaningful agency while improving performance by methodically mapping these roles to various task types based on current empirical findings. This framework lays the foundation for practically effective and morally sound human-AI collaboration that unleashes human potential by aligning task attributes to AI capabilities. It also provides structured guidance for context-sensitive automation that complements human strengths rather than replacing human judgment.

## 1. Introduction: A Gap in Current Human-AI Integration Approaches

The incorporation of artificial intelligence (AI) into human workflows offers considerable advantages; however, recent findings indicate a more complex situation. A thorough meta-analysis of 106 experimental studies has shown that the collaboration between humans and AI often yields results that are inferior to the best performances of either humans or AI when functioning independently [1]. This result calls into question the prevalent belief that the integration of human and artificial intelligence will always produce enhanced outcomes. Although human-AI systems typically surpass the performance of humans working in isolation, they frequently do not achieve the desired synergy, resulting in performance that is less effective than that of either humans or AI at their optimal levels [2],[3].

The disparity in performance is not consistent across various contexts. Vaccaro et al. (2024) found that the type of task plays a crucial role in moderating these results: the human-AI combinations generally performed worse than the best of humans or AI alone (Hedges' $g = -0.23$; 95% CI, -0.39 to -0.07) [1]. Task type emerged as a significant moderator ($F_{1,104} = 7.84$, $p = 0.006$), with decision-making tasks showing significant performance losses ($g = -0.27$; 95% CI, -0.44 to -0.10), while creation tasks demonstrated performance gains. Also, the relative ability of humans versus AI represented an even stronger moderator ($F_{1,104} = 81.79$, $p < 0.001$)—when humans outperformed AI independently, collaboration yielded significant performance gains ($g = 0.46$; 95% CI, 0.28 to 0.66), but when AI outperformed humans independently, collaboration resulted in significant performance losses ($g = -0.54$; 95% CI, -0.71 to -0.37). We illustrate these findings in Figure 1. These contrasting outcomes hold across different contexts, demonstrating that effective human-AI collaboration depends critically on both the type of task being performed and the comparative abilities of each party when working independently.

The healthcare sector exemplifies these intricate trade-offs. Dai and Singh (2023) investigated optimal strategies for AI implementation in healthcare pathways, concluding that the role of AI should vary according to patient risk profiles [4]. For low-risk patients, employing AI as a gatekeeper (initially screening patients) minimizes unnecessary treatments; for high-risk patients, utilizing AI as a secondary opinion (following specialist evaluation) reduces the likelihood of missed diagnoses. Notably, for intermediate-risk patients characterized by significant uncertainty, completely avoiding AI sometimes yielded superior outcomes—challenging the notion that AI is most beneficial in high-uncertainty situations [4],[5]. Rastogi et al. (2023) also characterize similar concerns, focusing on decision-making contexts. They offer a taxonomy of mechanisms through which human and ML-based decision-making may differ, potentially leading to complementary performance [6].

Beyond performance metrics, effective human-AI collaboration necessitates addressing human requirements for agency and control. He et al. (2023) identified some task dimensions that influence workers' preferences regarding automation: process consequence (the perceived cost of failure), social consequence (the threat of AI displacing humans), task familiarity, and task complexity [7]. These factors significantly affect the circumstances under which workers choose to delegate authority to AI assistants.

Considering these insights, we suggest a paradigm shift in our understanding of human-AI integration. Instead of fo-cusing on how humans can adjust to AI capabilities, we propose a task-oriented approach that aligns AI functions with the intrinsic characteristics of tasks. The central premise is: Task characteristics and how they fit the technological capabilities should determine appropriate AI roles, not the other way around. The remainder of the paper furthers this perspective and provides a detailed methodology and framework to do so.

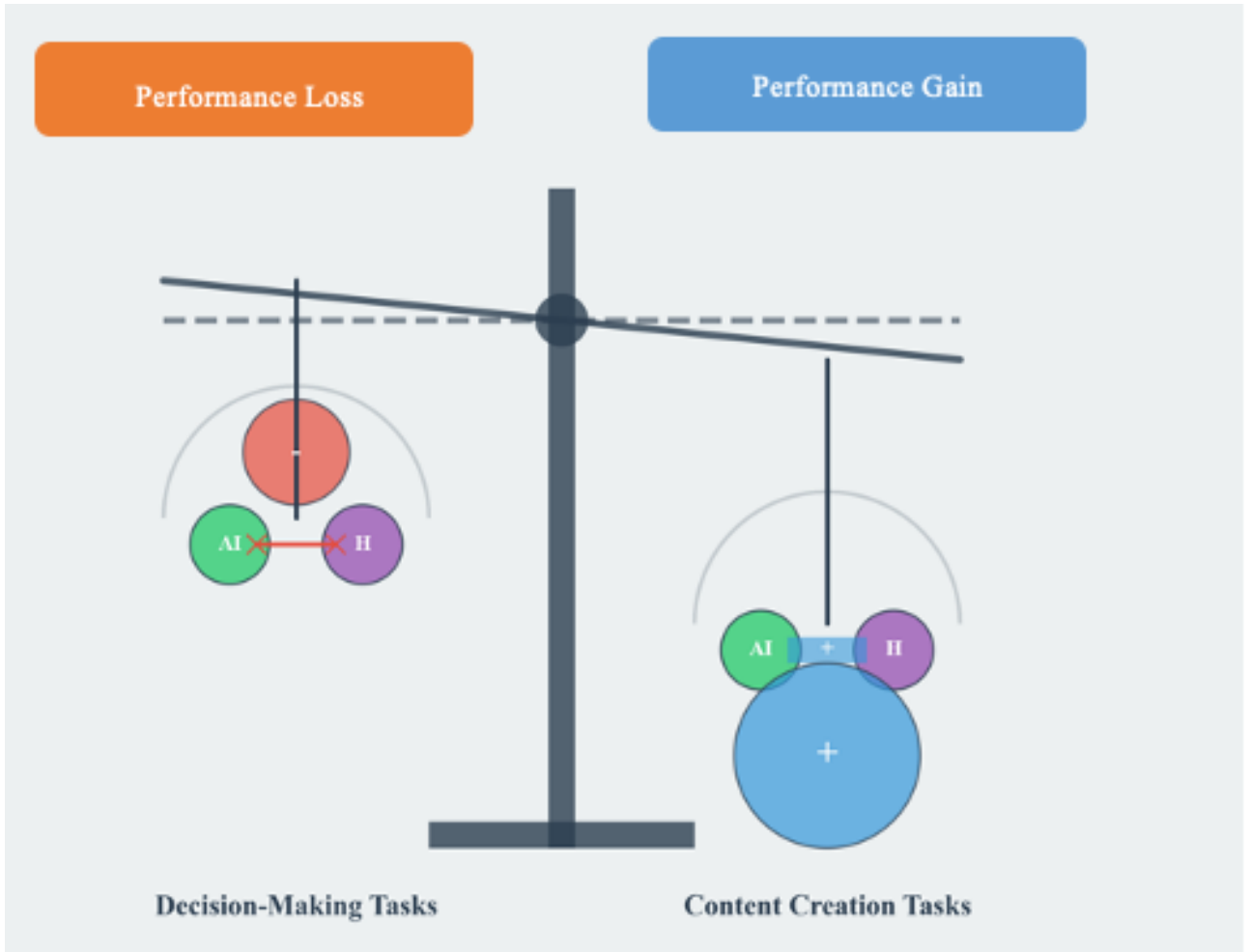

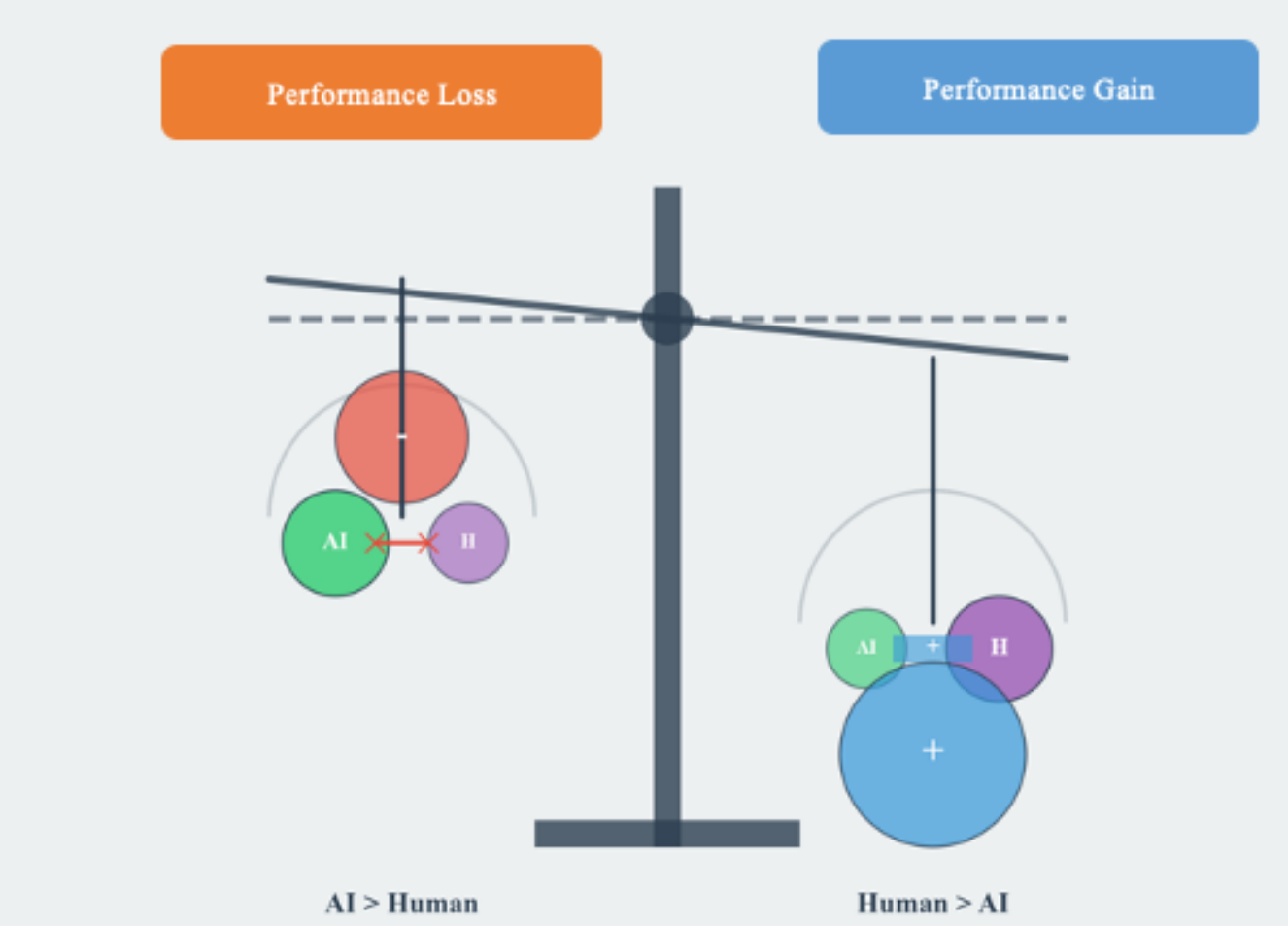

Figure 1: Impact of task type and relative ability on human-AI collaboration. The balance scale model illustrates how task type (left panel) and relative ability (right panel) affect human-AI collaboration performance. The balance scale metaphor visualizes how combined performance shifts toward gains (blue) or losses (red) depending on these two critical moderating factors.

## 2. The Core Dimensions of An Effective and Optimized Framework

An efficient and optimized interaction between humans and AI should be oriented towards specific tasks that encompasses various dimensions to inform suitable collaboration frameworks. This task-oriented model evaluates human-AI partnerships through two interrelated categories: task characteristics and AI roles.

In analyzing tasks, four critical factors must be taken into account. Firstly, the risk level (low, intermediate, high) indicates the potential repercussions of errors or failures, aligning with what He et al. (2023) referred to as 'process consequence'—the perceived cost associated with an AI assistant's mistakes. Secondly, complexity (simple, moderate, dynamic) pertains to the inherent difficulty, ambiguity, and cognitive demands of the task, which has been shown to significantly affect user preferences in interacting with AI systems [7]. Thirdly, stakeholder preferences: In addition to objective task metrics, the values of stakeholders play a crucial role in the effective deployment of AI. User preferences often encompass subjective elements such as the desire for human interaction, cultural traditions, perceptions of authenticity, or familiarity with the task, which may take precedence over mere efficiency in shaping suitable human-AI collaboration models. Lastly, the task type (decision-making, retrieval, action, creativity) delineates the fundamental objective of the activity. Vaccaro et al. (2024) highlighted this aspect as particularly significant, indicating that human-AI teams tend to perform less effectively in decision-making tasks while achieving better outcomes in creative tasks [1].

These AI roles characteristics subsequently inform the selection of the most suitable AI role among three primary options: *Autonomous AI*, which functions with minimal human oversight; *Assistive/Collaborative* AI, which augments human capabilities while ensuring substantial human involvement; and *Adversarial AI*, which offers critical counterarguments and alternative viewpoints to enhance decision quality and mitigate confirmation bias.

### 2.1 The Objectives of a Task-Driven Approach

This task-driven approach offers distinct advantages over prevalent integration methods. One key benefit is enabling dynamic adaptation between humans and AI systems. Instead of permanently allocating specific tasks to either humans or machines, this approach allows relationships to evolve naturally as users become more comfortable with AI assistance. As individuals become more accustomed to automated systems, their trust levels progressively influence the extent to which they delegate responsibilities to AI assistants in collaborative tasks.

Additionally, the framework's ability to facilitate a nuanced distribution of agency is noteworthy. By differentiating between initiative (the individual who initiates a task), control (the individual who oversees its execution), and de-cision-making (the individual who retains ultimate authority), organizations can strategically fine-tune human-AI collaboration. Lubors and Tan (2019) discuss how users typically preferred delegating initiative to AI for unfamiliar tasks while maintaining control and decision rights themselves [8].

The framework also ensures compatibility with natural human work patterns. For AI integration to succeed, it must complement how people naturally organize their activities. Research has highlighted several essential characteristics of efficient task management systems, such as distinct visual cues, the ability to prioritize tasks, and comprehensive filtering options that assist users in managing various simultaneous obligations [8], [9],[10].

This task-driven approach builds upon and extends previous research in meaningful ways. While Vaccaro et al. (2024) established task type as a significant factor in human-AI collaboration success, our framework provides a comprehensive methodology for categorizing tasks and matching them with suitable AI roles [1]. We expand on Dai and Singh's (2023) discovery that AI deployment should vary according to risk profiles, broadening this insight beyond healthcare to create a versatile framework applicable across industries [4]. Furthermore, we convert the findings of He et al. (2023) regarding the determinants of automation preferences into practical guidance for implementation [7].

This approach targets critical deficiencies in existing research concerning the identification of suitable human-AI collaboration frameworks tailored to specific task attributes. By employing a systematic analysis and providing practical illustrations from sectors such as healthcare, engineering, and creative industries, we demonstrate how well-aligned AI can improve productivity, enhance decision-making quality, and foster innovation. The framework guides practitioners in determining when AI should operate autonomously, collaboratively, or adversarially, based on objective task parameters and how the task characteristics fit the technological capabilities. This approach establishes a foundation for ethical AI deployment that amplifies human expertise rather than diminishing it.

## 3. Task Characteristics Across Risk, Complexity, and Stakeholders' Preference

We classify tasks based on their risk and complexity to identify suitable models for human-AI collaboration. This section provides an in-depth discussion of these dimensions, utilizing empirical findings from recent studies to create a detailed matrix of task characteristics. Recent meta-analyses in the field of human-centered AI research indicate that both risk and complexity play a crucial role in determining the effectiveness of human-AI collaboration. These dimensions not only impact performance results but also influence users' preferences regarding control, transparency, and modes of interaction.

Risk pertains to the potential repercussions of errors during task execution. Empirical evidence supports a three-tier categorization into low, intermediate, and high risk [7],[4]. Low risk tasks entail minimal negative outcomes and are easily reversible. Evidence suggests artificial intelligence can be effectively deployed as an operator for these tasks, since the potential harm from missing valid cases is relatively minimal [12],[13]. Intermediate risk tasks have moderate consequences, often involving sensitive information. Counterintuitively, it is discovered that for the intermediate risky and uncertain cases, "AI should be avoided altogether, neither as a gatekeeper nor as a second opinion" [4],[5]. High risk tasks have serious implications for human safety or involve sensitive information. Some studies suggest that faulty AI can mislead clinicians across all levels of expertise, including specialists, in critical medical situations such as cancer diagnosis and treatment. Therefore, even when AI shows superior average performance, involving humans in the process may be advantageous as they can help reduce the likelihood of rare but highly consequential errors [14],[15].

Complexity denotes cognitive demands and task ambiguity, which can also be classified into three levels: low, moderate, and high. Low complexity tasks are routine tasks with well-defined objectives and predictable results. Research indicates that conversational interfaces were more effective when participants had limited information to communicate [7]. Moderate complexity tasks require judgment based on multiple factors. As tasks become more intricate, users tend to prefer more comprehensive interfaces that enable them to access more functionality directly, allowing more efficient visibility and options [7]. High complexity tasks involve significant ambiguity and interdependent factors. Campero et al. (2022) emphasized that evaluation requires multiple criteria [16], while participants strongly preferred practical interfaces that grant more information and control compared to just conversational interfaces [17],[18].

### 3.1. Task Characteristics Matrix across Risk and Complexity

By mapping risk against complexity, we derive nine potential task categories that span the spectrum of human-AI collaboration scenarios (see Figure 2).

#### 3.1.1. Low Risk

While risk level determines the necessity of human oversight, complexity dictates the degree of AI support in low-risk contexts, from full automation for simple tasks to collaborative ideation for intricate work. Low risk, low complexity tasks are prime candidates for automation. He et al. (2023) discovered that participants expressed comfort with complete automation for specific tasks, emphasizing that any errors would likely be minor and easily rectifiable [7]. Similarly, Tyson et al. (2022) noted that routine activities are enhanced by automation when the potential for significant errors is minimal [19]. Low risk and moderate complexity tasks may exhibit some degree of uncertainty, yet their inherently low-risk characteristics permit partial automation with human oversight. For instance, AI can assist in basic content generation or standard report creation, although periodic validation may be necessary to maintain consistency. In scenarios characterized by low risk and high complexity, such as creative brainstorming or exploratory data analysis, the potential for errors is limited, despite the complexity involved. While AI can facilitate the ideation process, the role of human creativity is crucial for the refinement of the final outputs.

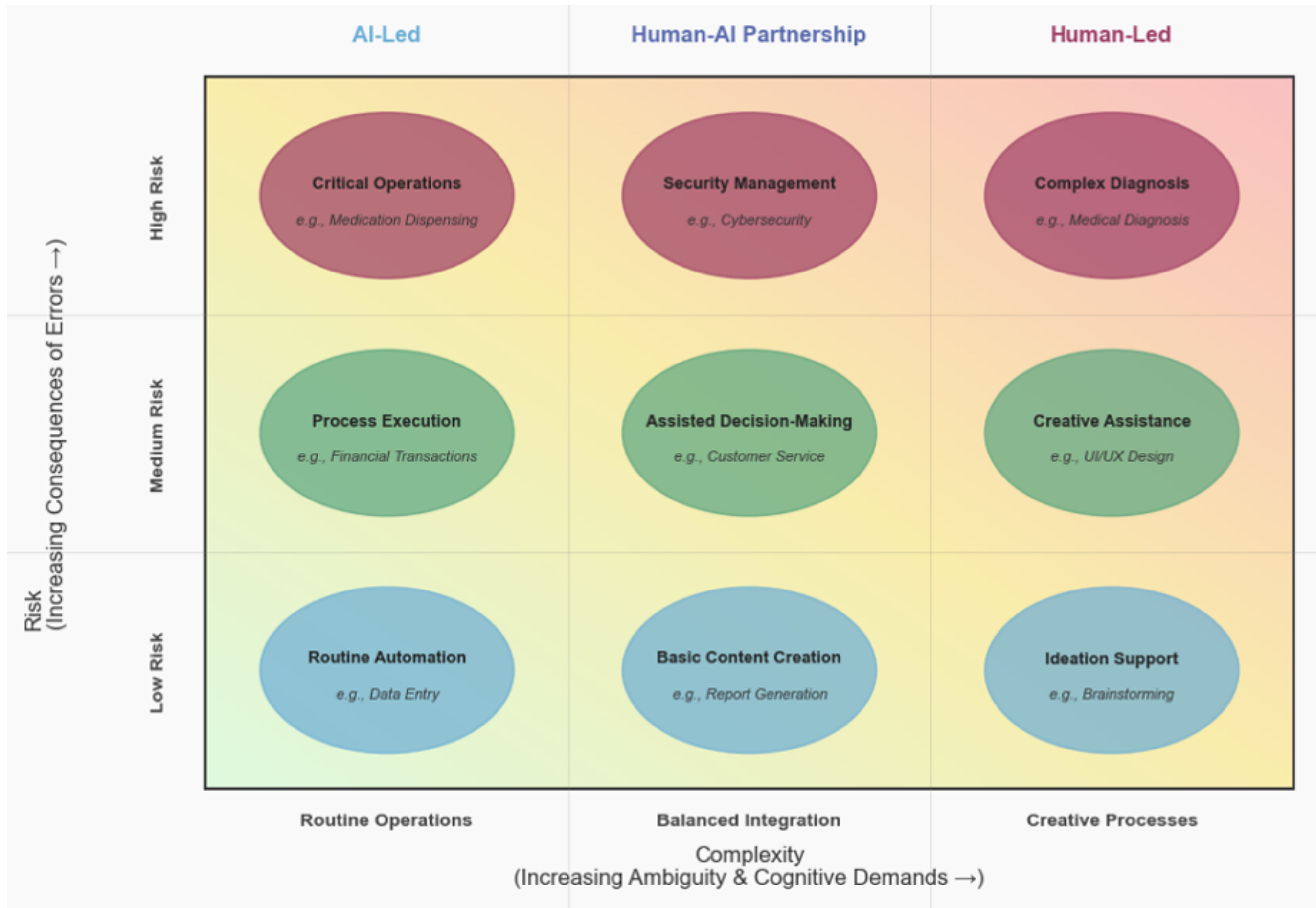

Figure 2: Task characteristics matrix (distribution of human-AI roles based on risk and complexity). This matrix illustrates how responsibilities for initiative shift from AI-led (left) to human-led (right) as complexity increases, and from autonomous operation (bottom) to heightened human oversight (top) as risk increases. Each cell represents a specific collaboration model with practical applications, demonstrating how different combinations of risk and complexity demand distinct approaches to human-AI integration that balance efficiency with appropriate levels of human agency and judgement.

### 3.1.2. Intermediate Risk

Tasks categorized as having intermediate risk and low complexity, such as basic financial transactions, can greatly benefit from AI-assisted workflows that incorporate built-in safeguards. The presence of human oversight is crucial for ensuring accuracy while utilizing automation to enhance efficiency. Intermediate risk, moderate complexity tasks require nuanced approaches. Compelling evidence have shown that for intermediate-risk scenarios, "a threshold $\tau GS$ exists" above which AI as a second opinion is preferable, and below which AI as a gatekeeper is optimal. Contrary to conventional assumption, they identified scenarios "where AI should not be used for intermediate-risk patients for whom uncertainty is highest" [4],[20]. When considering intermediate risk and high complexity tasks such as developing marketing strategies or designing user interfaces require diverse approaches and entail moderate consequences. A hybrid collaboration model, where AI proposes options and humans make the final decisions, may provide an effective balance between fostering innovation and mitigating risk.

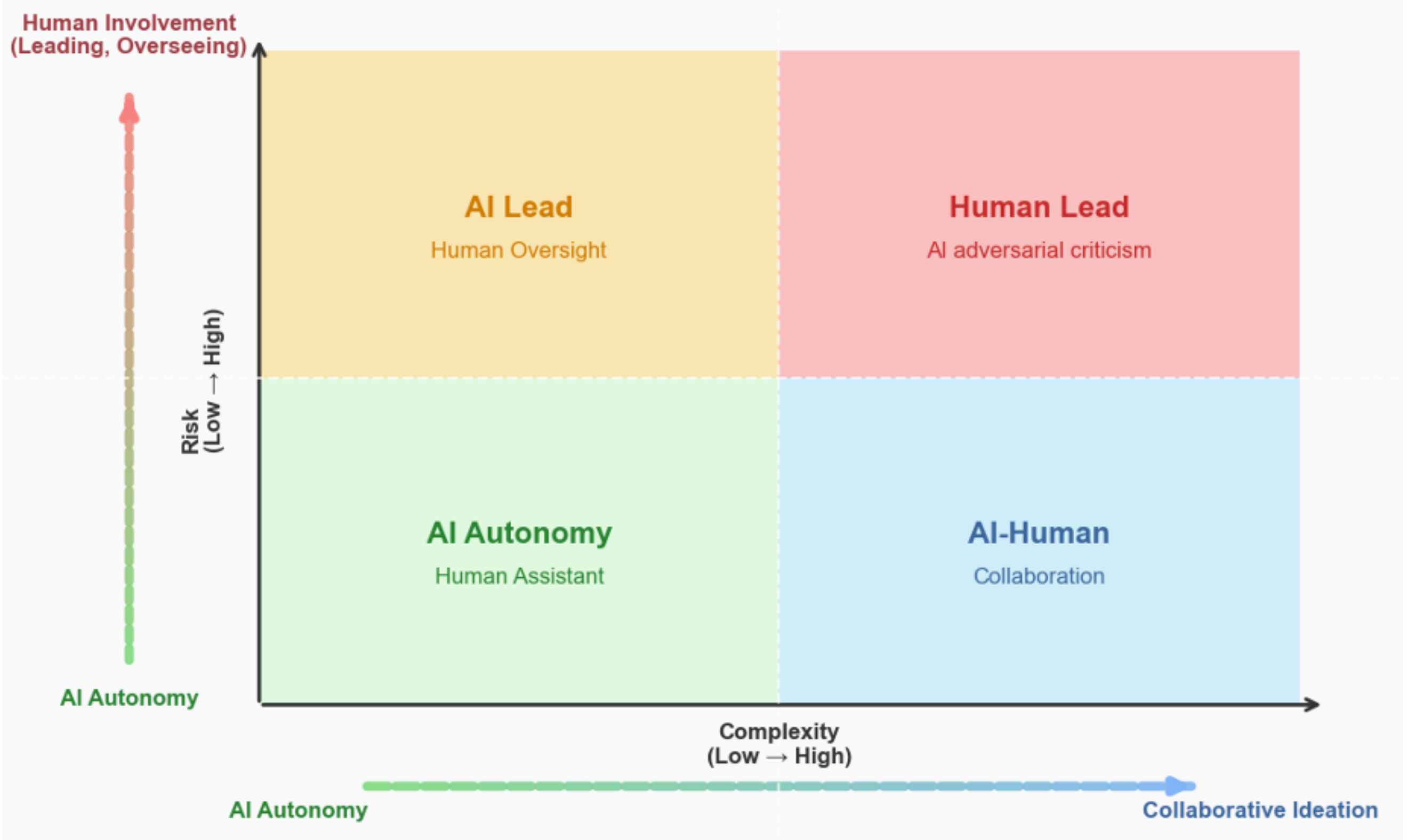

Figure 3: Human-AI collaboration dynamics: risk vs. complexity. Risk level determines the necessity of human involvement through leadership or oversight, while complexity dictates the degree of AI support. These two variables identify four approximately distinct zones: AI autonomy for simple, low-risk tasks; cooperative partnerships for complex, low-risk scenarios; human oversight for routine, high-risk operations; and human leadership with AI critique for intricate, high-risk challenges.

### 3.1.3. High Risk

In the context of high-risk tasks, regardless of their complexity, it is essential to implement meticulously calibrated human oversight. Here, the role of AI transitions from merely assisting in routine checks to engaging in complex decision-making processes, thereby ensuring accountability for critical outcomes. For high-risk tasks with low complexity, such as medication dispensing, stringent human oversight is imperative, even in cases of partial automation. Although errors are infrequent, their potential consequences are significant, necessitating robust fail-safes and verification protocols. High risk, moderate complexity tasks like financial auditing involve moderate complex processes but high stakes. AI can flag anomalies, but human judgment is critical for interpreting context and making final decisions.

High risk, high complexity tasks demand sophisticated human-AI collaboration. Several studies show that participants preferred having extra oversight and control to help guide these types of decisions [17],[18]. It is also emphasized that in high-stakes situations, maintaining human involvement in the process may still be necessary to minimize potentially costly errors [21]–[23].

### 3.1.4. Risk and Complexity Matrix

Taking all into account, as shown in Figure 3, the nuanced dynamics of optimal human-AI collaboration across varying risk and complexity profiles, revealing several critical insights. First, the framework suggests that in truly effective human-AI interaction, there exists no justification for complete human autonomy without AI involvement, as this would neglect a valuable resource that could enhance outcomes. Conversely, in low-risk, routine scenarios (bottom-left quadrant), full AI autonomy becomes advantageous, with humans potentially serving merely as assistants who just help and follow (not lead). This contrasts with the collaborative ideation zone (bottom-right), where humans and AI engage as genuine partners, iteratively building upon each other's contributions in complex but lower-risk contexts. As risk increases, human involvement necessarily intensifies, manifesting distinctly in the "oversight" mode (top left), where humans supervise AI operations without direct participation, versus the "adversarial criticism" mode (top right), where AI serves a crucial function in analyzing, challenging, and improving human-led decisions in high-stakes, complex scenarios. This adversarial criticism represents a sophisticated form of AI contribution, where the technology serves as a counterbalance to human cognitive biases and limitations rather than as a primary decision-maker. The gradient visualization reinforces that these interactions exist on a continuous spectrum, suggesting that organizations should strategically position their human-AI collaboration approaches based on precise risk-complexity assessments of specific tasks rather than implementing blanket policies that fail to account for these crucial contextual variations.

### 3.2. Task Characteristics Across Stakeholders' Preferences

Beyond risk and complexity, a third critical dimension is at work: stakeholder preference or task significance. This dimension acknowledges that subjective human values may supersede objective efficiency metrics when determining optimal AI roles. For example, parents might choose to read bedtime stories to their children rather than utilizing AI alternatives, emphasizing the importance of intimacy and connection over mere convenience. In a similar vein, within healthcare settings, patients often appreciate the human touch and emotional support provided by nurses and doctors, even when AI systems could deliver comparable or even superior diagnostic accuracy (see Table 1).

Furthermore, various psychological factors in marketing, such as the perception of authenticity, adherence to tradition, or preference for more natural methods, drive stakeholders to favor non-AI solutions for tasks that could be executed more swiftly or accurately by automated technologies. Familiarity with tasks also influences preferences for automation; users typically seek greater transparency and assistance for unfamiliar tasks but tend to favor less system support as their proficiency increases [24],[25].

It is crucial to recognize that stakeholder preferences are reciprocal, affecting both end users (patients, clients, consumers) and service providers (healthcare professionals, knowledge workers, creative practitioners). This aspect underscores the vital reality that effective human-AI collaboration must consider subjective value assessments alongside objective performance indicators, especially in scenarios where relationship-building, emotional support, or tradition hold substantial cultural significance.

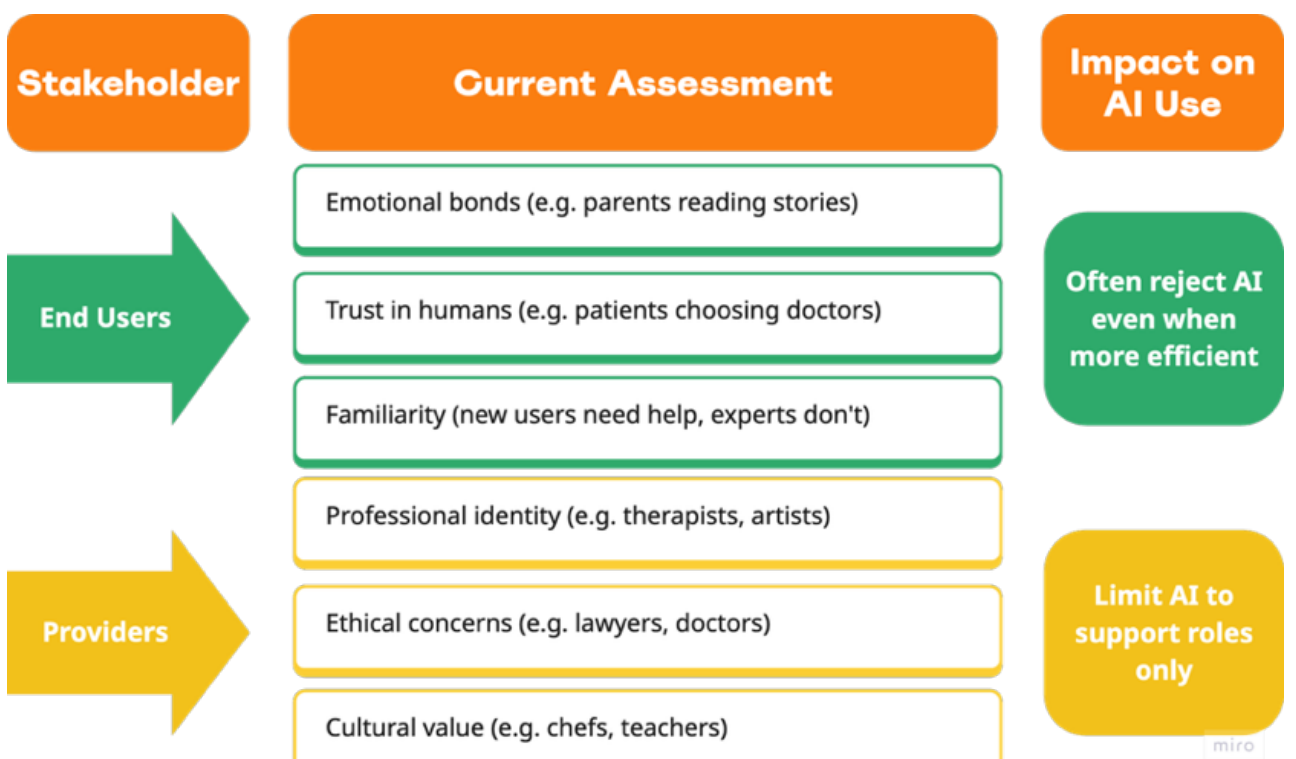

Table 1: Stakeholder AI preference factors.

## 4. Aligning AI Roles with Task Characteristics

Successful integration of AI systems into organizational workflows hinges on strategic role assignment based on inherent task suitability. This section investigates the efficacy of human-AI collaborations that leverage the unique strengths of both parties, facilitating an intelligent allocation of subtasks between human and artificial agents based on their distinct capabilities. We analyze this framework by pinpointing particular AI functions and their flexible adjustment across diverse operational environments [21].

### 4.1. Task Characteristics, AI Roles, and Optimal Applications

We identify three distinct AI roles, each suited to different task characteristics: autonomous AI, assistive/collaborative AI, and adversarial AI. Autonomous AI operates with minimal human intervention and, for instance, is optimal for low risk, low complexity tasks. Key characteristics include full task execution with minimal intervention, efficiency-optimized workflows, limited transparency, and high scalability [19],[20]. Assistive/collaborative AI works alongside humans for some contexts such as intermediate risk, moderate complexity tasks. This role leverages complementary strengths of both humans and AI, and effective systems need to skillfully assign subtasks to whichever partner—human or AI—has the greater capability in that specific area [1],[2],[21]. Essential features encompass collaborative execution with distinct human decision-making junctures, a harmonious balance of efficiency and quality, moderate levels of transparency, and the ability to adapt to diverse user requirements. Adversarial AI poses challenges to human assumptions, particularly in scenarios characterized by high risk and complexity. In situations requiring critical decision-making, human involvement can be beneficial, especially in mitigating the occurrence of rare yet significantly adverse outcomes [22],[23].

When handling cases with high risk factors, an approach that includes secondary verification (e.g. second-opinion model in medical issues) is more appropriate, particularly for patients where preventing diagnostic oversights is critical. Key characteristics include challenging assumptions, prioritizing robustness over efficiency, high transparency, and focus on critical edge cases (see Table 2).

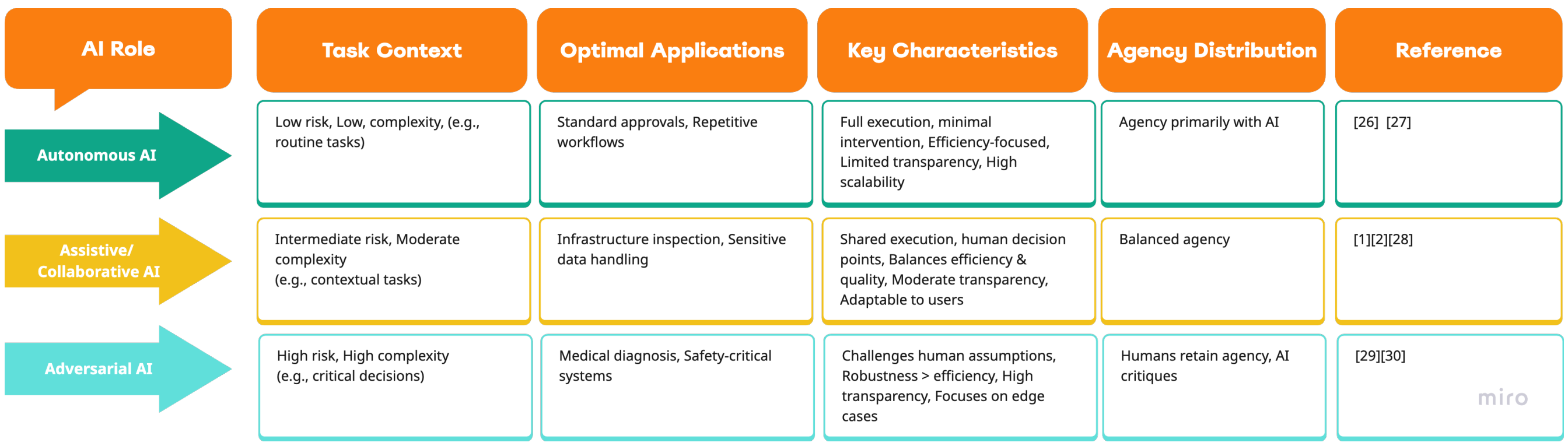

| AI Role | Task Context | Optimal Applications | Key Characteristics | Agency Distribution | Reference |
|---|---|---|---|---|---|
| Autonomous AI | Low risk, Low, complexity, (e.g., routine tasks) | Standard approvals, Repetitive workflows | Full execution, minimal intervention, Efficiency-focused, Limited transparency, High scalability | Agency primarily with AI | [26] [27] |
| Assistive/ Collaborative AI | Intermediate risk, Moderate complexity (e.g., contextual tasks) | Infrastructure inspection, Sensitive data handling | Shared execution, human decision points, Balances efficiency & quality, Moderate transparency, Adaptable to users | Balanced agency | [1][2][28] |
| Adversarial AI | High risk, High complexity (e.g., critical decisions) | Medical diagnosis, Safety-critical systems | Challenges human assumptions, Robustness > efficiency, High transparency, Focuses on edge cases | Humans retain agency, AI critiques | [29][30] |

Table 2: AI roles, task characteristics, and applications

### 4.2. Evidence-Based Agency Distribution and Dynamic Adaptation

The distribution of agency—initiative, control, and decision-making — varies across AI roles. For autonomous AI, most agency transfers to the system [19],[20]; for assistive AI, agency is balanced; for adversarial AI, humans retain primary agency while AI challenges assumptions, and importantly, optimal AI roles evolve dynamically [22],[23]. The task-driven framework emphasizes adaptation based on evolving expertise, changing risk profiles, shifting task complexity, and developing trust.

This task-oriented methodology responds to the findings that collaborations between humans and AI frequently fall short due to discrepancies between the requirements of tasks and the frameworks of collaboration [1],[2],[4]. By deliberately taking into account the attributes of tasks, organizations can transcend the mere optimization of automation, fostering genuinely synergistic partnerships that not only augment human abilities but also uphold dignity and autonomy [23].

## 5. Operationalizing the Framework Across Task Categories

This section demonstrates the implementation of the task-driven framework across various tasks, highlighting the impact of risk and complexity factors on the effective roles of AI in practical applications (see Table 3).

### 5.1. Decision-Making Tasks

Inappropriate AI roles can hinder decision-making processes, ranging from excessive automation of critical tasks to insufficient use of AI in everyday operations. Here, we show how this framework's task-centered approach resolves these gaps in information retrieval and critical judgments. a. High Risk, High Complexity Decision-Making in Medical Diagnosis: Medical diagnosis exemplifies high-risk, high-complexity decision tasks where the task-driven framework recommends adversarial AI. This aligns with the idea of getting a second medical opinion, which is particularly valuable for high-risk patients where avoiding missed diagnoses is the top priority [26]. This approach reduces missed diagnoses by approximately 38%-68% compared to no AI [4].

Additional research corroborates this methodology, emphasizing that successful collaboration between humans and AI necessitates the system's overall proficiency in effectively distributing subtasks to the most suitable partner [1],[27]. In the context of b. low-risk, low-complexity decision-making for routine approvals, such as standard time-off requests, the task-driven framework advocates for the use of autonomous AI. This recommendation is reinforced by findings indicating that participants expressed comfort with complete automation for tasks where potential errors would be relatively minor and easily rectifiable. Furthermore, participants showed a preference for conversational interfaces in straightforward workflows, enabling them to initiate automation with minimal effort [28].

### 5.2. Information Retrieval Tasks

Information retrieval spans from low-risk, low-complexity tasks (answering FAQs) to high-risk, moderate-complexity ones (sensitive legal research). For routine inquiries, autonomous AI is appropriate; and for sensitive information retrieval, assistive/collaborative AI is recommended, as participants wanted significant oversight when handling sensitive information, concerned about whether "that information is secure, if it's going anywhere after that.

### 5.3. Action Implementation

The process of action implementation presents distinct challenges, as neither complete automation nor solely human involvement is ideal. The task-oriented framework illustrates how assistive AI improves results in areas such as infrastructure inspection, and document creation by effectively integrating human insight with AI functionalities. The task of infrastructure inspection is characterized by intermediate risk and moderate complexity, leading the task-oriented framework to advocate for the use of assistive or collaborative AI. This strategy merges AI's ability to recognize patterns with human oversight, which corresponds with findings indicating that participants preferred more advanced interfaces for tasks of moderate complexity, seeking traditional graphical user interfaces for enhanced control [7]. Document creation may entail high complexity but is generally associated with low risk. The task-oriented framework recommends employing AI in a supportive or collaborative capacity, a suggestion bolstered by studies showing that teams comprising both humans and AI achieved significantly superior outcomes compared to either humans or AI systems operating in isolation [1],[29],[7].

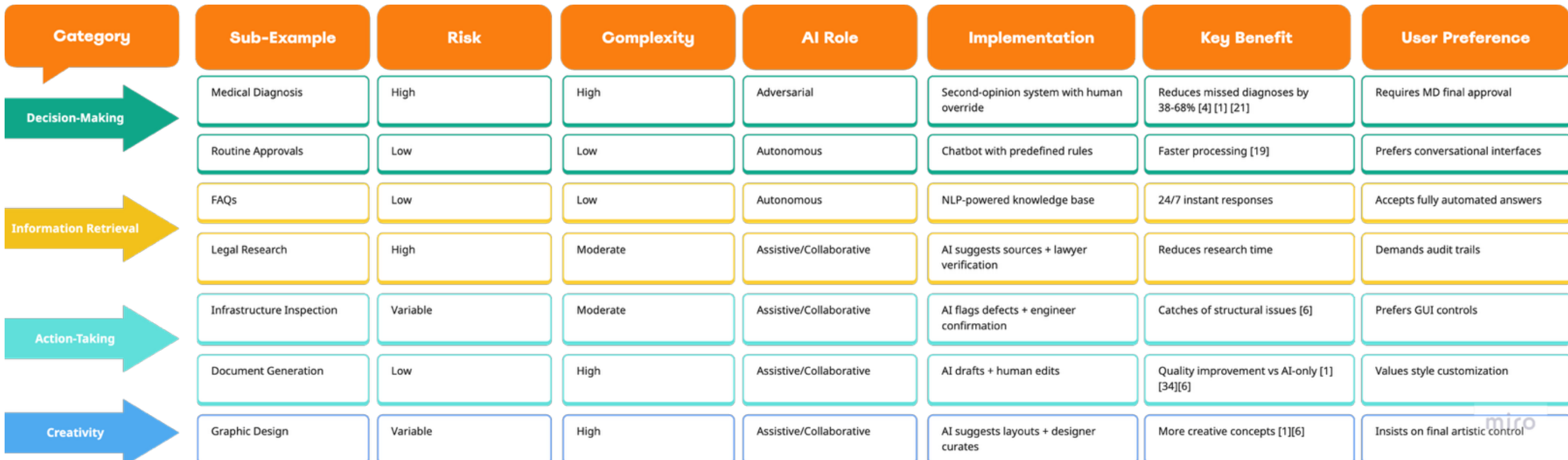

| Category | Sub-Example | Risk | Complexity | AI Role | Implementation | Key Benefit | User Preference |
| --- | --- | --- | --- | --- | --- | --- | --- |
| Decision-Making | Medical Diagnosis | High | High | Adversarial | Second-opinion system with human override | Reduces missed diagnoses by 38-68% [4] [1] [21] | Requires MD final approval |
| | Routine Approvals | Low | Low | Autonomous | Chatbot with predefined rules | Faster processing [19] | Prefers conversational interfaces |
| Information Retrieval | FAQs | Low | Low | Autonomous | NLP-powered knowledge base | 24/7 instant responses | Accepts fully automated answers |
| | Legal Research | High | Moderate | Assistive/Collaborative | AI suggests sources + lawyer verification | Reduces research time | Demands audit trails |
| Action-Taking | Infrastructure Inspection | Variable | Moderate | Assistive/Collaborative | AI flags defects + engineer confirmation | Catches of structural issues [6] | Prefers GUI controls |
| | Document Generation | Low | High | Assistive/Collaborative | AI drafts + human edits | Quality improvement vs AI-only [1] [34][6] | Values style customization |
| Creativity | Graphic Design | Variable | High | Assistive/Collaborative | AI suggests layouts + designer curates | More creative concepts [1][6] | Insists on final artistic control |

Table 3: Framework applications by task category.

### 5.4. Creativity Tasks

Creative design tasks are marked by high complexity and fluctuating risk levels. The task-oriented framework consistently promotes the use of assistive or collaborative AI, a stance supported by research that highlights a significantly more favorable impact of human-AI collaboration in creative tasks as opposed to decision-making scenarios. This is consistent with the finding that participants wished to maintain control and decision-making authority for tasks necessitating subjective judgment, even while appreciating AI support [7],[1].

## 6. Task-Driven Approach and Human Agency and Dignity

While performance optimization dominates much discourse around human-AI collaboration, preserving human agency and dignity is equally critical. Several distinct types of agency in human-AI collaboration are identified, including: initiative (who starts a task), control (who decides how the task gets done), and decision-making (who has final authority) [35],[36]. The distribution of these factors significantly affects human dignity within professional environments. As the risk associated with tasks escalates, the necessity for substantial human oversight correspondingly intensifies.

### 6.1. Balancing Efficiency and Agency Across Risk Levels

The task-oriented framework confronts the essential conflict between utilizing AI for efficiency and maintaining human agency, contingent upon varying levels of task risk. In scenarios characterized by high risk and complexity, it is advisable to implement AI systems that bolster human cognitive processes while safeguarding human decision-making authority, thereby optimizing agency in high-stakes situations.

Conversely, as risk diminishes, a greater delegation of tasks to AI becomes justifiable for the sake of efficiency, although mechanisms for human oversight remain crucial to uphold a suitable balance of agency. Evidence suggests that users prefer to retain control over system functionalities even in low-risk contexts, particularly concerning communication parameters, highlighting that concerns regarding agency persist irrespective of the risk involved. Notably, the assistance provided by AI does not inherently lead to improved outcomes, despite potential accuracy benefits [20]. For patients classified as medium risk with highly uncertain conditions, it is imperative to prioritize human agency by refraining from AI utilization altogether. This recommendation is consistent with research indicating that combinations of human and AI efforts can sometimes yield inferior results compared to their independent operations. These findings reinforce our methodology for assessing when and how AI participation genuinely adds value while honoring the essential equilibrium between efficiency and human agency [2],[38].

### 6.2. Social Dimensions and Professional Identity

Human-AI collaboration has important social dimensions affecting workplace dignity. Social consequences, including reputation management and impression management, constitute another dimension of tasks along which information needs and desired control vary. Participants were particularly concerned about having AI represent them to others, especially to important clients or teammates. The task-oriented framework addresses this issue by advocating for increased human oversight in tasks that carry significant social implications. A fundamental tenet is that artificial intelligence should augment, not supplant, human knowledge. This perspective aligns with the notion that human-AI collaboration is most effective when human performance exceeds that of AI alone, indicating that AI serves best as an enhancement to human skills. Successful collaboration necessitates customizing the role of AI according to both the performance of the AI system and the expertise of the specialist. When professionals retain adequate control over communications that could affect their professional relationships or reputation, the collaborative framework upholds dignity while optimizing the advantages of technological support. These insights imply the necessity of designing AI systems that provide varying levels of control based on social implications, ensure transparency regarding contributions, complement specialized knowledge, and adjust autonomy levels in accordance with the social aspects of tasks [34].

### 6.3. Three Key Ethical Safeguards

The task-oriented framework establishes essential ethical safeguards that uphold human autonomy in AI-enhanced settings. Initially, it introduces strategic oversight mechanisms for critical decisions through adversarial AI systems that question assumptions while honoring human decision-making authority—a characteristic that participants in He et al. (2023)'s research specifically requested for significant choices. Furthermore, it intentionally assigns humans to roles that necessitate nuanced judgment and creativity, utilizing AI as an auxiliary tool rather than a substitute, in recognition of users' expressed wish to maintain control over subjective evaluations. Lastly, it mitigates the risk of excessive reliance on AI by pinpointing specific situations where algorithmic intervention could be more detrimental than beneficial, especially in intermediate-risk medical contexts where uncertainty is heightened and human clinical judgment prevails. These protective measures collectively guarantee that the integration of AI enhances rather than undermines human agency in professional settings.

### 6.4. Practical Implementation Through Design

Effectively translating these principles into operational systems necessitates careful interface design strategies derived from empirical studies. Preview functionalities serve as a vital component, enabling users to review and assess automated processes prior to execution, thus preserving significant oversight. Detailed control features that visually represent varying levels of risk facilitate appropriate human intervention according to the complexity of tasks. Adaptive support systems that modify assistance based on user proficiency promote skill enhancement while mitigating excessive dependence on automation. Ultimately, versatile interface designs customized for various task demands guarantee optimal collaboration between humans and AI across a range of contexts. Collectively, these design aspects foster work environments where technology enhances human capabilities instead of supplanting critical judgment, achieving a delicate equilibrium between efficiency and substantial human input in an increasingly automated professional landscape.

## 7. Conclusion and Future Directions

This study presents a dynamic framework centered on task-driven human-AI collaboration, emphasizing the importance of task characteristics over technological capabilities. By categorizing tasks according to their risk and complexity and aligning them with suitable AI roles, this framework offers a systematic method for designing collaborations that enhance human abilities while maintaining agency and dignity.

This research contributes significantly to the domain of human-AI collaboration in several ways. Firstly, it introduces a task-driven perspective that highlights the critical role of task type in influencing collaboration effectiveness. Secondly, it provides a detailed mapping of task characteristics to AI roles—whether autonomous, assistive/collaborative, or adversarial—illustrating the necessity for varied approaches based on risk profiles. Thirdly, it underscores the importance of preserving human agency and dignity through a thoughtful distribution of initiative, control, and decision-making, acknowledging the desire of workers to be co-creators in their technological environments. Lastly, it contests the common belief that greater AI involvement invariably leads to better outcomes, demonstrating that in scenarios characterized by intermediate risk and high uncertainty, human judgment may still be superior.

The task-driven framework also carries practical implications for organizations seeking to implement human-AI collaboration. It necessitates domain-specific calibration, urging organizations to evaluate their distinct task profiles while considering additional factors such as social implications and user familiarity. Furthermore, it advocates for the integration of dynamic adaptation mechanisms to accommodate the evolving nature of task characteristics and user needs, especially as users become more acquainted with tasks and technologies. Finally, it suggests that interface design should correspond with task complexity, as evidence indicates that users favor different interaction modalities based on task requirements. It is crucial for organizations to strike a balance between optimizing performance and preserving significant human agency, recognizing that upholding human dignity and purpose may necessitate accepting certain performance compromises.

Although the task-driven framework offers a systematic method for human-AI collaboration, it is not without its significant limitations. Future investigations should seek to validate this framework within specific domains and contexts, assessing its applicability across diverse professional settings. Furthermore, additional task dimensions, such as temporal factors, social interactions, and emotional elements, deserve further exploration. The development of genuinely adaptive systems that can respond to evolving task characteristics is another promising avenue for research. Lastly, longitudinal studies that track the evolution of agency preferences over time would yield important insights into the fluid dynamics of human-AI interactions.

By embracing a task-oriented approach, organizations can develop AI-enhanced environments that not only operate efficiently but also uphold human dignity and purpose. The genuine promise of human-AI collaboration is found not in the substitution of human labor, but in the augmentation of human abilities, fostering systems where technology enhances rather than undermines the uniquely human contributions that are vital in our progressively automated society.

**Acknowledgement**: This research was funded in part by the SUNY-IBM AI Research Alliance. Furthermore, in accordance with MLA (Modern Language Association) guidelines, we acknowledge the use of AI-powered tools, such as Anthropic's applications, for their assistance and support including in writing, editing, and improving the clarity and organization of the manuscript.